\RequirePackage[2020-02-02]{latexrelease}
\documentclass[twocolumn,prl,amsmath,amssymb,showpacs,superscriptaddress,floatfix]{revtex4}
\usepackage{graphicx}% Include figure files
\usepackage{bm}% bold math
\usepackage{lineno,hyperref}
\usepackage{epsf}
\usepackage{xcolor}
\usepackage[english,russian]{babel}
\begin{document}
%\draft
\title{Collective excitations in liquid carbon tetrachloride: a molecular dynamics study}%:
%First-Order versus Continuous Transition }

\author{Yu. D. Fomin \footnote{Corresponding author: fomin314@mail.ru}}
\affiliation{Vereshchagin Institute of High Pressure Physics,
Russian Academy of Sciences, Kaluzhskoe shosse, 14, Troitsk,
Moscow, 108840, Russia } 

\author{V. V. Brazhkin}
\affiliation{Vereshchagin Institute of High Pressure Physics,
Russian Academy of Sciences, Kaluzhskoe shosse, 14, Troitsk,
Moscow, 108840, Russia } 
\date{\today}

\begin{abstract}

We perform a molecular dynamic study of collective excitations of carbon tetrachloride and compare the results with
experimental data from the literature. The data of simulations are in good argeement with the experimental ones. 
The results of the simulations confirm the presence of large positive sound dispersion (PSD) in carbon tetrachloride,
which should be related to some relaxation processes which do not take place in atomic systems.

{\bf Keywords}:speed of sound
\end{abstract}

\pacs{61.20.Gy, 61.20.Ne, 64.60.Kw}

\maketitle

%\begin{keywords}
%  boiling line, near-critical maxima (Widom line), phase equilibria of mixtures
%\end{keywords}

\section{Introduction}

Collective excitations are of increadible importance for description of different properties of crystals. Although it is not so well recognized in the literature,
the collective excitations are of the same importance for liquids too. A model which describes the thermodynamic properties of liquids, such as heat capacity,
was proposed recently \cite{ph-theory}. Importantly, this model successfully describes the isochoric heat capacity of liquids of different nature: rare gases, liquid metals, 
molecular liquid, etc.

The collective fluctuations of density of a liquid are described by the dynamic structure factor $S({\bf k},\omega)$, where {\bf k} is the wave vector and $\omega$ is the frequency
 \cite{hansen}. In hydrodynamic limit, i.e., when a liquid can be considered as a
continuum medium, the dynamic structure factor consists of a central peak and two symmetric side peaks at $\omega = \pm c_sk$, where $c_s$ is the adiabatic speed of sound and \cite{boonyip}. 
However, at small $k$ when the liquid cannot be represented as a continuum medium a Positive Sound Dispersion (PSD) is often observed, 
i.e. the measured excitation frequence exceeds the hydrodynamic value: $\omega>c_s k$.

The development of experimental techniques, such as X-ray scattering, allowed to perform direct measurements the dynamic structure factors of liquids. The spectra of collective excitations
of liquids extracted from these data. Several well recognized works on spectra of liquid metals were published in the literature, including such systems as liquid iron, copper, zinc \cite{hos-1},
tin \cite{hos-sn,hos-sn-1}, sodium \cite{hos-na}, gallium \cite{hos-ga,mok-ga} and some other metals. Some semiconductors were also probed (see \cite{hos-si}
for the spectra of silicon). A powerfull theoretical method of calculation of the dynamical structure factor is reviewed in \cite{tmf}. This approach was 
successfully employed to describe the collective excitations in a number of system like, for example, liquid gallium \cite{mok-ga}, lithium \cite{m-li}, alluminium \cite{m-al}, et. al.

Collective excitations were also measured in a number of molecular liquids, for instance, water \cite{wat-1,wat-2}. Interestingly, as it was found from molecular dynamics simulation, 
the dependence of the excitation frequency of water on thermodynamic parameters shows anomalous behavior, i.e., behaves unlike the one of other liquids \cite{w1,w2,w3}. For instance,
while in normal fluid the frequency at fixed wave vector increases upon isochoric heating, in water in can decrease. In the case of isotherms excitation frequency at given $\bf{k}$ shows
strong dependence on the pressure (or density), while in the case of water there is a region where the frequency does not change with pressure (see \cite{wat-2} for experimental results and
\cite{w1} for molecular dynamics simulation). Simulation also shows anomalous behavior along isobars, however, up to now there is no experimental confirmation, since it requires
rather high temperature, above the one in reported experiments \cite{wat-1}.

As it follows from the experimental and theoretical investigations, liquids demonstrate PSD when they are close to the melting line. However, the range of PSD (i.e., 
$(\omega(k)-c_s \cdot k)/(c_s \cdot k)$) can be rather different for different liquids. PSD in liquid metals is usually of the order of several percent (up to $10-12 \%$ \cite{scop}). 
PSD becomes very large in the vicinity of structural transformations in a liquid, like it is observed in water, which can be described by a mixtrure of two different local structures (
see, for instance, a review \cite{tale} and references therein).
 For example, PSD in water is about $110 \%$  \cite{vvb-jept}. 
Tellurium, which resembles many unusual properties of water, also has PSD about $70 \%$ \cite{hos-te}. 
An extremely large PSD of $150 \%$ is reported in liquid mercury in the vicinity of its critical point \cite{vvb-hg}.

Interestingly, large PSD is observed in molecular liquid even if no structural transformations is observed. Large PSD was observed in carbon tetrachloride 
($30 \%$ PSD is reported in \cite{ccl4-1} and $50 \%$ in \cite{ccl4-2}), benzene (about $50 \%$ \cite{benz-psd}) and aceton
(up to $65 \%$ \cite{aceton-psd}). Although this effect is observed in a set of completely different liquids, the results strongly depend on the technical details. For instance,
the works \cite{ccl4-1} and \cite{ccl4-2} were done by the same group of authors, but different methods of data manupulation were used. While in the former work the
experimental spectra were fitted to Damped Harmonic Oscillator (DHO) function, Generalized Langevin analysis \cite{gla} was used in the later, which resulted in much larger
PSD. However, in any case there is unambigously high PSD in molecular systems, which has no explanation up to now.

In the present paper we try to find out the reasons of extremely large PSD in the simples molecular liquid mentioned above: carbon tetrachloride. The molecule $CCl_4$ is
highly symmetric, it has neither dipole, nor quadrupole moments, but it demonstrates large PSD. We perform molecular dynamics simulations and analyze the results to find
the origin of this large PSD.

%It is seen that the effect of very large PSD is experimentally observed in a large number different molecular liquids, i.e. it should be rather general.
%At the same time there is no any satisfactory explanation of this phenemenon.

%%%%%%%%%%%%%%%%%%%%%%%%%%
%In our previous works we have shown that two regimes of microscopic dynamics of fluids are possible: solid-like rigid fluid and gas-like fluid, which are 
%separated by so called Frenkel line \cite{ufn-fr,pre-fr,jept-fr,scf-fr}. The Frenkel line in (P,T) or ($\rho$,T) plane is defined as a line where transverse excitations of a
%liquid disappear \cite{jpcm-2018}. Another interesting phenomena which takes place in rigid fluid, but not in a gas-like one, is Positive Sound Dispersion (PSD). PSD means that
%the frequency of excitation at some wave vector $\bf{k}$ exceeds the hydrodynamic values $c_s^{hyd}=c_s \cdot  k$, where $c_s$ is the hydrodynamic speed of 
%sound. Positive sound dispersion also disappears at the Frenkel line \cite{iron,antibryk,neon}.

%Although the concept of Frenkel line was proposed basing  on the results of molecular simulations, it rapidly received an experimental verifications \cite{neon}.
%A recent review gives a good description of the current situation in experimental investigation of the Frenkel line \cite{fr-exp}.
%%%%%%%%%%%%%%%%%%%%%%%%%%%%%%%

\section{System and Methods}

In the present work we perform molecular dynamics simulation of a system of 2000 molecules of $CCl_4$ in a cubic box with periodic boundary conditions.
The interactions are modeled within OPLS-AA force field \cite{opls-1,opls-2,opls-3}. The molecules were considered as rigid bodies, i.e. all $C-Cl$ bonds
and $Cl-C-Cl$ angles were fixed. 

Firstly the system was simulated for 5 ns at constant temperature $T=300$ K and constant pressure $P=1$ bar in order to find the equilibrium density. Then
the system was equilibrated for more 5 ns at the found density (constant volume simulation). The time step at these two steps was 1 fs. Finally we simulated the system
in microcanonical ensemble (constant number of particles N, volume V and energy E) for more 5 ns to calculated the time correlation functions. 

The wave vector dependent velocity current is defined as $J({\bf k}, t)=\sum_{j=1}^N {\bf v}_j exp^{-i {\bf kr}_j}$. The longitudinal and transverse correlation
functions of the velocity current are defined as:

\begin{equation}
   C_L({\bf k},t)=\frac{k^2}{N}<J_z({\bf k},t)J_z({\bf -k},0)>,
\end{equation}
and

\begin{equation}
   C_T({\bf k},t)=\frac{k^2}{2N}<J_x({\bf k},t)J_x({\bf -k},0)+J_y({\bf k},t)J_y({\bf -k},0)>,
\end{equation}
respectively \cite{gla}. The frequency of excitation at given wave vector ${\bf k}$ is given as the location of the peak of Fourier transform of the corresponding correlation function.

The speed of sound is calculated as $c_s=\gamma^{1/2} c_T$, where $c_T=\left( \frac{\partial P}{\partial \rho} \right)_T$, $\gamma=c_P/c_V$ is the ratio of heat capacities.
All derivatives are calculated from numerical simulation of the molecular dynamics data.

We additionally calculate the velocity autocorrelation function (VACF) and take its Fourier transform to get the Rahman frequency of the system. Radial distribution
function (RDF) of the centres of mass of the molecules is used to characterize the structure of the system.

All calculations were performed with LAMMPS simulation package \cite{lammps}.

\section{Results and Discussion}

The density of carbon tetrachloride obtained in simulation is $\rho=1.569$ $g/cm^3$. The experimental density is $\rho_{exp}=1.59$ $g/cm^3$, i.e. the relative 
error is about $1.3 \%$, which is reasonable.

Figure 1 (a) and (b) show the RDF and static structure factors of center of mass (or  carbon atoms) of $CCl_4$ molecules. It is seen that these plots
correspond to low-temperature liquid. RDF demonstrates three well defined peaks, i.e., the system is a well-structured liquid. We compare the static structure factor
from MD simulations with the one from neutron measurements of Ref. \cite{ccl4-sk}. It is seen that there is a good agreement between the simulation
and the experiment for the first peak of $S(k)$. The agreement is also reasonable for the second peak of $S(k)$. The deviation between the simulation
and experiment increases with increasing of the value of the wave vector.

%The second peak of both $g(r)$ and $S(k)$ have non-gaussian shape, therefore
%although the molecule is highly symmertic the structure of the liquid cannot be properly fitted by the one of some simple system, such as Lennard-Jones one.

\begin{figure}

\includegraphics[width=8cm, height=6cm]{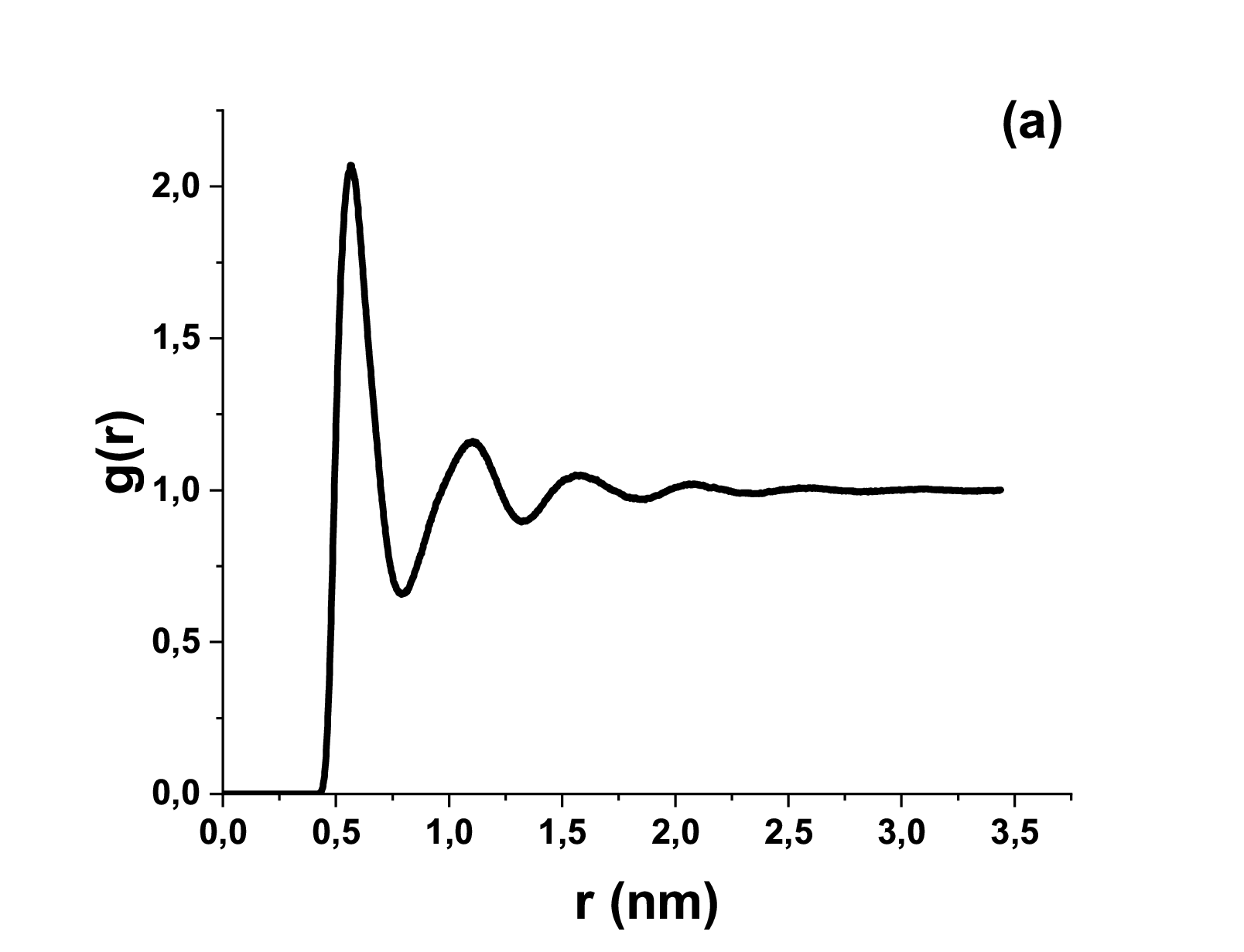}%

\includegraphics[width=8cm, height=6cm]{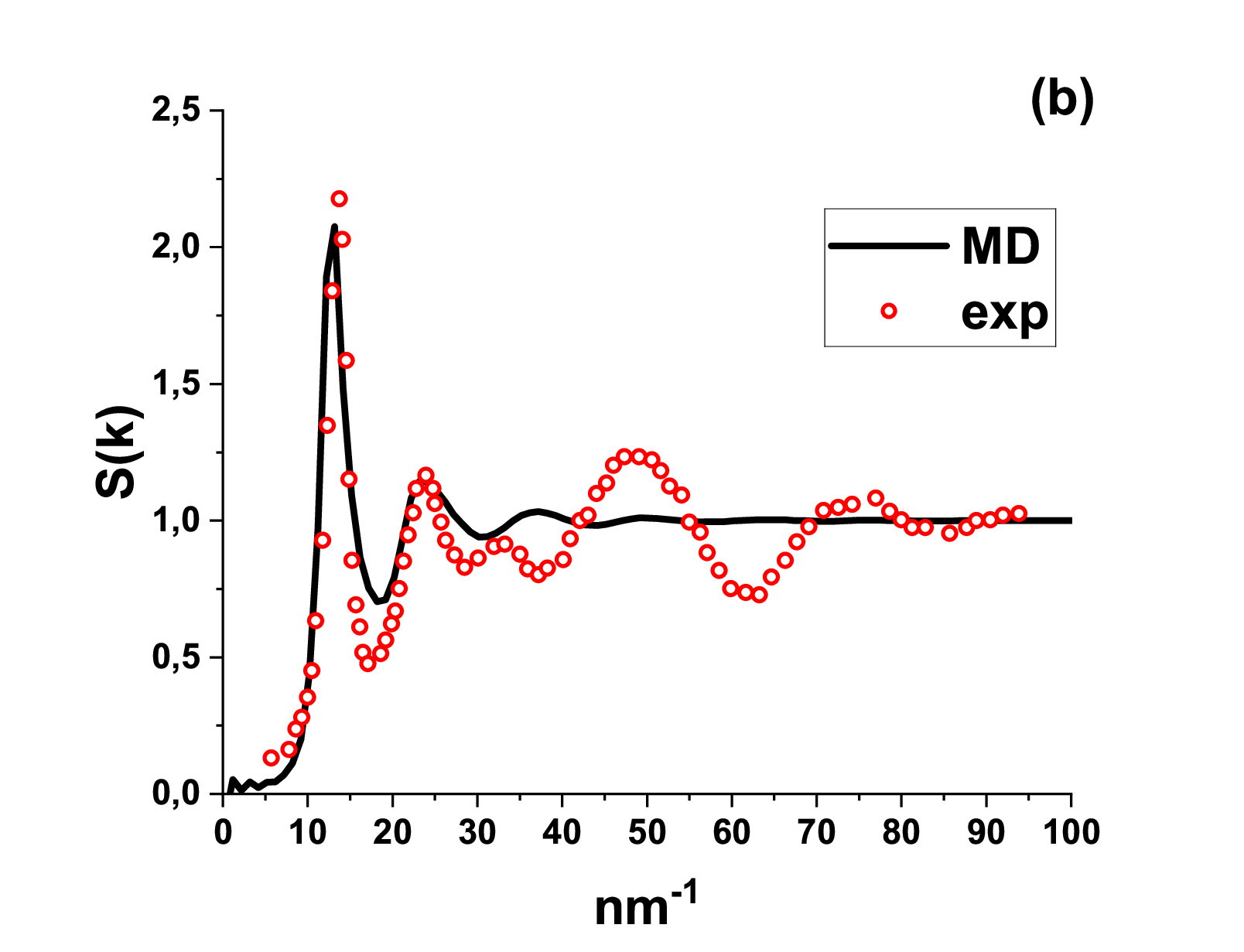}%

\caption{\label{rdf} (a) Radial distribution function and (b) static structure factor of centers of mass of carbon tetrachloride molecules at $T=300$ K and ambient pressure.
The experimental data are taken from Ref. \cite{ccl4-sk}}
\end{figure}

Figure 2 shows the excitation spectra of carbon tetrachloride obtained in the present work. Experimental data from Ref. \cite{ccl4-2} are shown for a comparison. 
The calculated hydrodynamic speed of sound is $c_s=980.12$ $m/s$, while the experimental one is $c_s^{exp}=921.5$ $m/s$, i.e. the relative difference 
$((c_s-c_s^{exp})/c_s^{exp}=6.4 \%$. The relative difference of between the excitation frequencies from simulation and from experiment are of the 
same order of magnitude, but the frequencies from simulation are lower than the experimental ones. Therefore, the molecular simulation strongly underestimates
the PSD in $CCl_4$. Moreoever, the underestimation is due to both possible factors: underestimation of the frequency and overestimation of the hydrodynamic speed of sound.
However, the maximum range of PSD obtained in simulation is $\frac{\omega(k)-c_sk}{c_sk} \approx 30 \%$, i.e. it is still much larger than the one of atomic systems, but
lower than the results of the recent experiments.

\begin{figure}

\includegraphics[width=8cm, height=6cm]{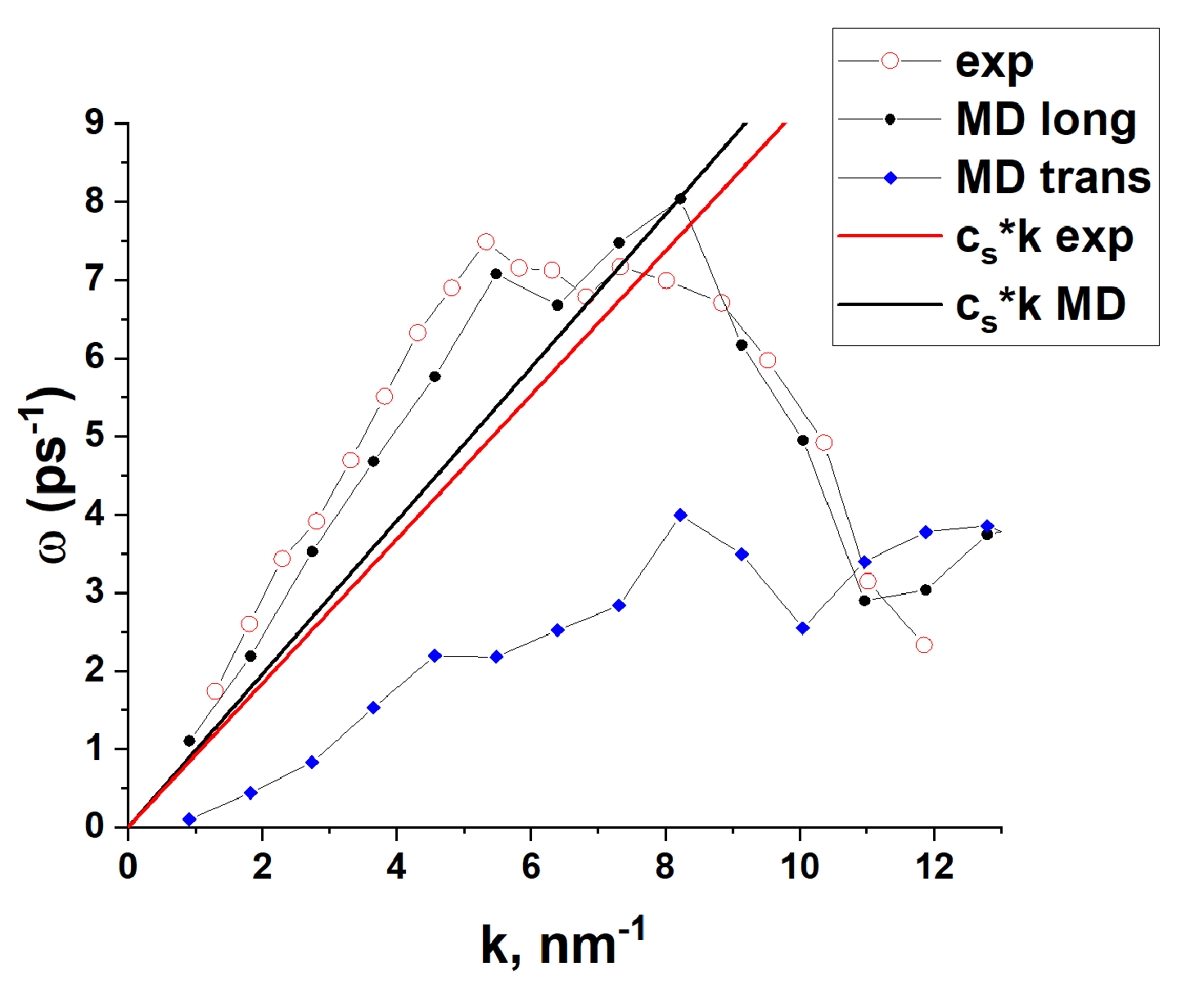}%

\caption{\label{sp} Spectra of longitudinal and transverse excitations of carbon tetrachoride at $T=300$ K and $P=1$ bar. The experimental data are taken from \cite{ccl4-2}.}
\end{figure}

The molecules of cacron tetrachloride $CCl_4$ are very symmetric. Therefore, one might expect that they should be well described by some simple spherically-symmetric potential,
like a Lennard-Jones one. At the same time it is seen from Fig. 1 that the second peaks of both RDF and structure factor have a shoulder at low $r$ in case of $g(r)$ and
high $k$ in the case of $S(k)$, which is not typical for simple liquid. The critical parameters of $CCl_4$ are $T_c \approx 556$ K, $P_c=45$ bar and $\rho_c \approx 0.557$ $g/cm^3$.
Therefore, the conditions of the present study are $T/T_c=0.54$ and $\rho / \rho_c =2.817$. In the case of LJ system ($T_c^{LJ}=1.321$, $\rho_c^{LJ}=0.316$ in the dimensionless
units based on the LJ potential parameters) the point with the same reduced temperature and density is $T_{LJ}=0.713$ and $\rho_{LJ}=0.89$. This point is located in
a solid-liquid two phase region of the LJ system, i.e., an atomic system with such thermodynamic conditions starts to crystallize, which the molecular $CCl_4$ is in a liquid state.

LJ parameters of carbon tetrachloride obtained from the viscosity data are given in the book \cite{bood-lg}. The parameters are $\varepsilon_{LJ}=322.7$ K and
$\sigma_{LJ}=5.947$ \AA. Using these parameters one can calculate that the reduced density and temperature of $CCl_4$ in the present study
are $\rho ^*=1.29$ and $T^*=0.93$ which is in the crystalline region of the phase diagram of LJ system. 

It is seen that in spite of a simple symmetric shape of the molecule carbon tetrachloride cannot be effectively represented by a simple LJ-like system. The possibility
of approximation of molecular liquid by a simple LJ-like models was discussed in \cite{vvb-ufn}. It was shown that LJ model can be sufficiently good
to describe the properties of crystalline molecular systems, such as binding energy or bulk modulus. However, as soon as the molecules start to move from
the crystalline cites, the shape of the molecule becomes of principle importance. For this reason, simple isotropic models fail to describe the dynamical
properties of molecular liquids, such as diffusion coefficient of shear viscosity. Our results show that an isotropic potential cannot describe the dynamic structure
factor even of a system with symmetric molecules. As mentioned above, the LJ parameters of $CCl_4$ obtained from the viscosity measurements bring to the 
fact that carbon tetrachloride at ambient conditions should be solid. This contradiction is in a good agreement with our conclusions that simple isotropic models
are inapplicable to the dynamic properties of molecular liquids.

 Moreover,
in the present study we employ the simplest model which consist of rigid molecules, i.e., no internal degrees of freedom can be responsible for the high
PSD of carbon tetrachloride. At the same time we observe that the microscopic dynamics of carbon tetrachloride is strongly different from the one of LJ-like systems. Figure 3 
shows a VACF of the carbon atoms which coincide with the VACf of centers of mass and the geometrical centers of the molecules. While in a simple
liquid VACF usually demonstrates one or two well define oscillations and then fluctuates around zero, here we observe that VACF rapidly decays slightly below
zero and then slowly relaxes. Therefore, some slow processes take place in the microscopic dynamics of the system which are not observed in simple
atomic liquids.

\begin{figure}

\includegraphics[width=8cm, height=6cm]{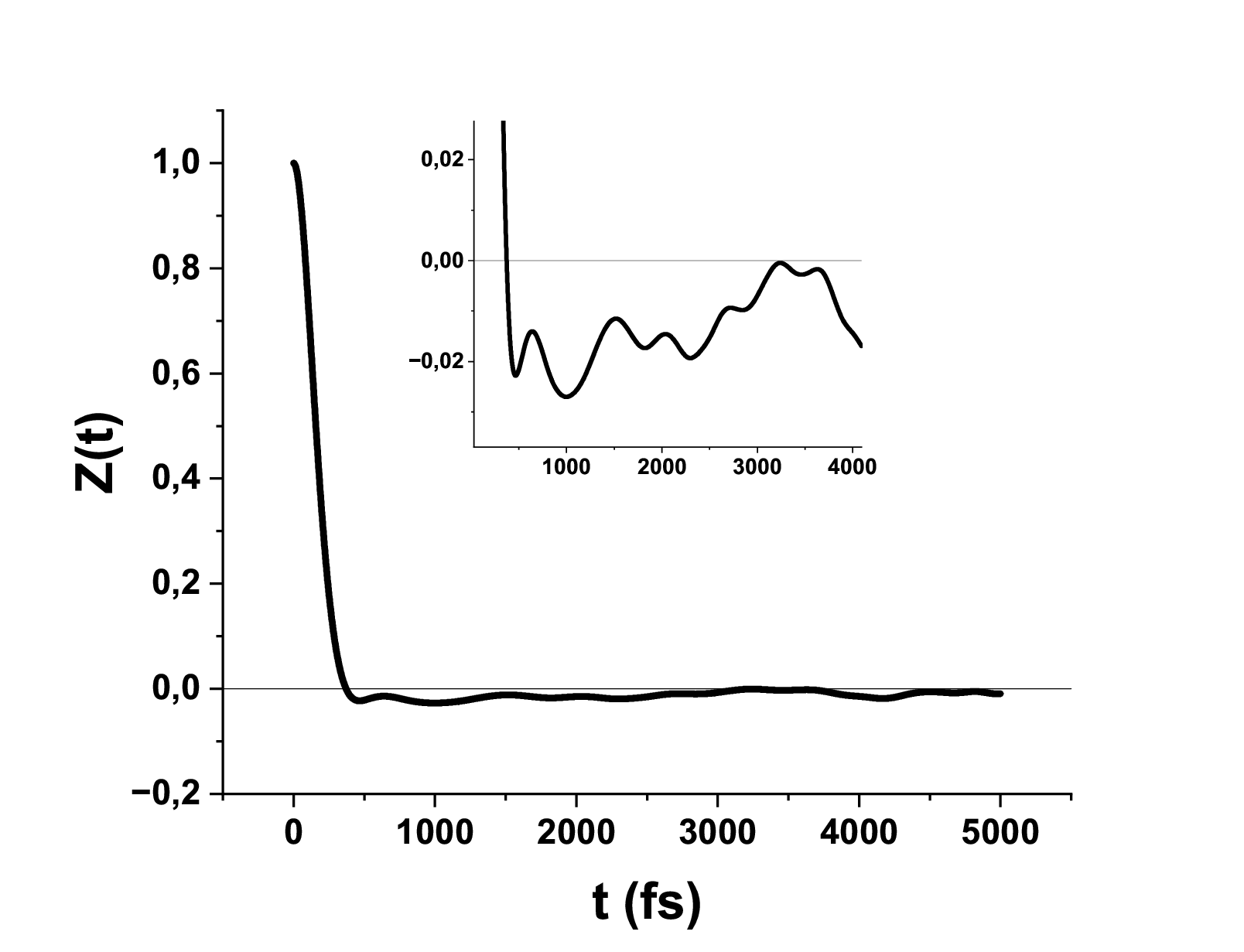}%

\caption{\label{vacf} Velocity autocorrelation function of carbon atoms of $CCl_4$ molecules at $T=300$ K and $P=1$ bar.}
\end{figure}

As it was mentioned in Introduction, large PSD appears when some structural transformations in the liquid take place. It leads to a strong frequency dependence of 
the bulk modulus $B(\omega)$ \cite{vvb-jept}. No structural changes are found in carbon tetrachloride at ambient conditions, therefore other mechanisms should be responsible 
for large PSD in this system. At the same time more relaxation processes are expected in molecular systems which should be responsible for the strong frequency 
dependence of the bulk modulus.

\section{Conclusions}

A molecular dynamics study of excitation spectra of carbon tetrachloride is perfromed. The results are compares with existing experimental data.
Large PSD (about $30 \%$) is found in the system, which should be related to the presence of some slow dynamic processes, which 
are identified from the velocity autocorrelation function of the system. It is also shown that despite of the simple symmetric 
shape of the molecules carbon tetrachlodire cannot be represented by a simple Lennard-Jones-like model.

This work was carried out using computing resources of the federal
collective usage center "Complex for simulation and data
processing for mega-science facilities" at NRC "Kurchatov
Institute", http://ckp.nrcki.ru, and supercomputers at Joint
Supercomputer Center of the Russian Academy of Sciences (JSCC
RAS). The calculation of Widom lines was supported by Russian Science Foundation (Grant 19-12-00111).
The calculation of boiling lines was supported by Russian Science Foundation (Grant  19-12-00092).


\begin{thebibliography}{62}




\bibitem{ph-theory} K. Trachenko, V. V. Brazhkin, Collective modes and thermodynamics of the liquid state, 
Rep. Prog. Phys. 79(1), 016502 (2016) DOI: 10.1088/0034-4885/79/1/016502 

\bibitem{hansen} Hansen J P, McDonald I R Theory of Simple Liquids (London:
Academic Press, 1986)

\bibitem{boonyip} J. P. Boon and S. Yip, Molecular Hydrodynamics (McGraw-Hill
International Book Company, New York, 1980).

\bibitem{hos-1} S. Hosokawa, M. Inui, Y. Kajihara, S. Tsutsui and A. Q. R. Baron, Transverse excitations in liquid Fe, Cu and Zn,
J. Phys.: Cond. Matt.: 27, 194104 (2015)

\bibitem{hos-sn} S. Hosokawaa, J. Greif, F. Demmel, W.-C. Pilgrim, Quasielastic lineshape of liquid Sn studied
by inelastic X-ray scattering, Chemical Physics 292, 253–261 (2003)

\bibitem{hos-sn-1} S. Hosokawa,  S. Munejiri, M. Inui, et. al., Transverse excitations in liquid Sn, J. Phys.: Condens. Matter 25 (2013) 112101

\bibitem{hos-na} W.-C. Pilgrim, S. Hosokawa, H. Saggau, H. Sinn, E. Burkel, Temperature dependence of collective modes in liquid sodium,
Journal of Non-Crystalline Solids 250-252, 96-101 (1999)

\bibitem{hos-ga} S. Hosokawa, M. Inui, Y. Kajihara, Transverse Acoustic Excitations in Liquid Ga, Phys. Rev. Lett. 102, 105502 (2009).

\bibitem{mok-ga} A. V. Mokshin, R. M. Khusnutdinoff, A. G. Novikov, N. M. Blagoveshchenskii, and A. V. Puchkov, Short-Range Order and Dynamics
of Atoms in Liquid Gallium, J. Exp. and Theor. Phys., 121, 828–843 (2015)

\bibitem{m-li} A V Mokshin and B N Galimzyanov, . Phys.: Condens. Matter 30 (2018) 085102 (17pp)

\bibitem{m-al} A V Mokshin, R M Yulmetyev, R M Khusnutdinoff and P Hanggi, J. Phys.: Condens. Matter 19 (2007) 046209 (16pp) 

\bibitem{hos-si} S. Hosokawa, J. Greif,  F. Demmel, W.-C. Pilgrim, Phonon dynamics of liquid Si – inelastic X-ray
scattering studies, Nuclear Instruments and Methods in Physics Research B 199, 161–164 (2003)

\bibitem{tmf} A.V. Mokshin, R. M. Khusnutdinov, Ya. Z. Vilf, B. N. Galimzyanov, Theoret. and Math. Phys., 206:2 (2021), 216–235

\bibitem{iron} Yu D Fomin, V N Ryzhov and V V Brazhkin, J. Phys.: Condens. Matter 25 (2013) 285104.


\bibitem{wat-1} T. Yamaguchi, K. Yoshida, N. Yamamoto, S. Hosokawa, M. Inui, A.Q.R. Baron d, S. Tsutsui, J. Phys. and Chem. of Solids 66,  2246–2249 (2005)

\bibitem{wat-2} U. Ranieri, P. Giura, F. A. Gorelli, M. Santoro, St. Klotz, Ph. Gillet, L. Paolasini, M. Marek Koza, and L. E. Bove, J. Phys. Chem. B 120, 34, 9051–9059 (2016)

\bibitem{w1} Yu. D. Fomin, E. N. Tsiok, V. N. Ryzhov, V. V. Brazhkin, Journal of Molecular Liquids 287, 110992 (2019) 

\bibitem{w2} Yu. D. Fomin, E. N. Tsiok, V. N. Ryzhov, V. V. Brazhkin, Fluid Phase Equilibria 498, 45050 (2019)

\bibitem{w3} Yu. D. Fomin, E. N. Tsiok, V. N. Ryzhov, V. V. Brazhkin, Journal of Molecular Liquids 337, 116450 (2021)

\bibitem{vvb-jept} V. V. Brazhkin, I. V. Danilov, O. B. Tsiok, 
 JETP Letters, 117:11 (2023), 834–848

\bibitem{tale} P. Gallo, K. Amann-Winkel, Ch. Austen Angell, et. al., Chem. Rev. 116, 7463-7500 (2016)



\bibitem{hos-te} Y. Kajihara, M. Inui, S. Hosokawa, K. Matsuda, and A. Q. R. Baron, J. Phys.: Condens. Matter 20, 494244
(2008).

\bibitem{vvb-hg} V. V. Brazhkin, E. Bychkov, and O. B. Tsiok, Phys. Rev.
B 95, 054205 (2017)

\bibitem{scop} T. Scopigno and G. Ruocco, Rev. Mod. Phys. 77, 881-933 (2005)

\bibitem{ccl4-1}  T. Kamiyama, Sh. Hosokawa, A. Q. R. Baron, S.  Tsutsui, K. Yoshida, W.-Ch. Pilgrim, Yo. Kiyanagi, and T. Yamaguchi, 
J. Phys. Soc. Jpn. 73, 1615-1618 (2004)

\bibitem{ccl4-2} Sh. Hosokawa, K. Yoshida, J. Mol. Liq. 395, 123828 (2024)

\bibitem{benz-psd} Sh. Hosokawa, K. Yoshida, J. Mol. Liq. 390, 123063 (2023)

\bibitem{aceton-psd} Sh. Hosokawa, T. Kamiyama, K. Yoshida, T. Yamaguchi, S. Tsutsui, A. Q.R. Baron, J. Mol. Liq. 332, 115825 (2021)

\bibitem{gla} J. P. Boon and S. Yip, Molecular Hydrodynamics, McGraw-Hill, New York, 1980


%%%%%%%%%%%%%%%%%%%%%%%%%5

\bibitem{ufn-fr} V. V. Brazhkin, A. G. Lyapin, V. N. Ryzhov, K. Trachenko, Yu. D. Fomin, E. N. Tsiok, Physics ± Uspekhi 55 (11) 1061 - 1079 (2012)

\bibitem{pre-fr} V. V. Brazhkin, Yu. D. Fomin, A. G. Lyapin, V. N. Ryzhov, and K. Trachenko, Phys. Rev. E 85, 031203 (2012)

\bibitem{jept-fr} V. V. Brazhkin, Yu. D. Fomin, A. G. Lyapin, V. N. Ryzhov, and K. Trachenko, JETP Letters 95, 164–169 (2012)

\bibitem{scf-fr} V. V. Brazhkin, A. G. Lyapin, V. N. Ryzhov, K. Trachenko, Yu. D. Fomin, and E. N. Tsiok, Russian Journal of Physical Chemistry B 8, 1087–1094 (2014)

\bibitem{jpcm-2018} Yu. D. Fomin , V. N. Ryzhov, E. N. Tsiok, J. E. Proctor, C. Prescher, V. B. Prakapenka, K. Trachenko and V. V. Brazhkin, J. Phys.: Condens. Matter 30, 134003 (2018)

\bibitem{antibryk} Yu. D. Fomin, V. N. Ryzhov, E. N. Tsiok, V. V. Brazhkin and K. Trachenko, J. Phys.: Condens. Matter 28, 43LT01 (2016)

\bibitem{neon} C. Prescher, Yu. D. Fomin, V. B. Prakapenka, J. Stefanski, K. Trachenko, and V. V. Brazhkin, Phys. Rev. B 95, 134114 (2017)

\bibitem{fr-exp} C. Cockrell, V.V. Brazhkin, K. Trachenko, Physics Reports 941, 1–27 (2021)


%%%%%%%%%%%%%%%%%%%%%%%%%


\bibitem{opls-1} W. L. Jorgensen, J. Tirado-Rives, J. Am. Chem. Soc. 110 (6): 1657–1666 (1988)

\bibitem{opls-2} W. L. Jorgensen, D. S. Maxwell, J. Tirado-Rives, J. Am. Chem. Soc. 118 (45): 11225–11236 (1996)

\bibitem{opls-3} W. L. Jorgensen, J. Tirado-Rives, PNAS  102, 6665-6670 (2005)



\bibitem{lammps} A. Thompson et. al., Computer Physics Communications 271, 108171 (2022)

\bibitem{ccl4-sk} P. Jovari, Gy. Meszaros, L/ Pusztai, E. Svab Physica B 276-278, 491-492 (2000)


\bibitem{bood-lg} Bruce E. Poling, John M. Prausnitz, John Paul O'Connell "The Properties of Gases and Liquids", McGraw-Hill (2001), New York

\bibitem{vvb-ufn}  V. V. Brazhkin, “Can glassforming liquids be ‘simple’?”, Phys. Usp., 62:6 (2019), 623–629

\end{thebibliography}
\end{document}